# Vulnerability Analysis for Captchas Using Deep Learning

Jaskaran Singh Walia
School of computer science and engineering,
Vellore institute of technology, India
karanwalia2k3@gmail.com

Aryan odugoudar
College of Engineering Computing and Cybernetics (CECC),
Australian National University (ANU), Canberra, Australia
aryanodugoudar143@gmail.com

*Abstract*— Several websites improve their security and avoid dangerous Internet attacks by implementing CAPTCHAs (Completely Automated Public Turing test to tell Computers and Humans Apart), a type of verification to identify whether the end-user is human or a robot. The most prevalent type of CAPTCHA is text-based, designed to be easily recognized by humans while being unsolvable towards machines or robots. However, as deep learning technology progresses, development of convolutional neural network (CNN) models that predict text-based CAPTCHAs becomes easier. The purpose of this research is to investigate the flaws and vulnerabilities in the CAPTCHA generating systems in order to design more resilient CAPTCHAs. To achieve this, we created CapNet, a Convolutional Neural Network. The proposed platform can evaluate both numerical and alphanumerical CAPTCHAs.

## I. INTRODUCTION

The CAPTCHA (Completely Automated Public Turing Test to Tell Computers and Humans Apart) is a test used on the internet websites to distinguish humans from robots, harmful programs, or other computerized agents that attempt to impersonate human intelligence.

Therefore, a CAPTCHA is majorly used to prevent various types of cyber security threats, penetration, and attacks aimed at the anonymity of webapps, websites, webservices, user's credentials, or even in semiautonomous programs and driver assistance systems when an actual human is required to take control of a program/machine. These attacks, in particular, frequently result in circumstances in which programs try to replace humans and attempt automating services to send a large number of unsolicited texts/mails, access the databases, and influence online surveys or polls. DDOS attacks are one of the most frequent types of cyber-attacks, in which the target is overloaded with highly unexpected traffic, mostly to find the credentials of the target or to cripple the target system for a short period of time.

In the development of cybersecurity systems, one of the most classic yet highly successful methods are the use of a CAPTCHA system. As a result, the attacker machines can be identified, and odd traffic can be restricted or ignored to minimize harm. Hence research on testing the vulnerabilities of CAPTCHA images is essential because it aids in the detection of weak areas and finds loopholes in previously generated CAPTCHAs, which leads to the prevention of these flaws in newly designed CAPTCHA-generating systems, hence improving Internet security. Background noise, image distortion, rotation, text warping, variable string length, and character merging are just a few of the technologies that have been devised to make text-based CAPTCHAs more robust.

However, as an effect of the rapid advancement of deep learning algorithms recently, CAPTCHA recognizing systems are now more capable than ever of predicting most of the latest text-based CAPTCHA security measures. As a result, advanced security procedures must be created to strengthen these CAPTCHAs against malicious systems.

As deep-learning algorithms specialize in extracting relevant details from an image and have various applications in fields such as object detection. Deep learning technique's resilient properties make them perfect for developing robust image recognition networks to undertake attacks on CAPTCHAs. Although several such detection algorithms use digital image processing techniques in their implementation, these systems have limitations (e.g., low feature extraction and easily influenced by noise in the images).

Therefore, advanced deep-learning approaches are gradually replacing conventional strategies. Recent research methods follow character segmentation and then recognition. However, they cannot solve newer, more challenging CAPTCHAs with skewed letters that vertical lines cannot separate. As a result, rectangular windows are ineffective for segmentation, and more powerful classification algorithms are required.

CAPTCHAs can be in several forms, for example, numeric or alpha- numeric voice, strings or images. Fig.1 displays samples of alpha-numerical CAPTCHAs and their various types. In this paper we will focus mainly on alphanumeric based CAPTCHA since they are more commonly used in websites with high traffic and dense networks mainly due to their lower computational cost.

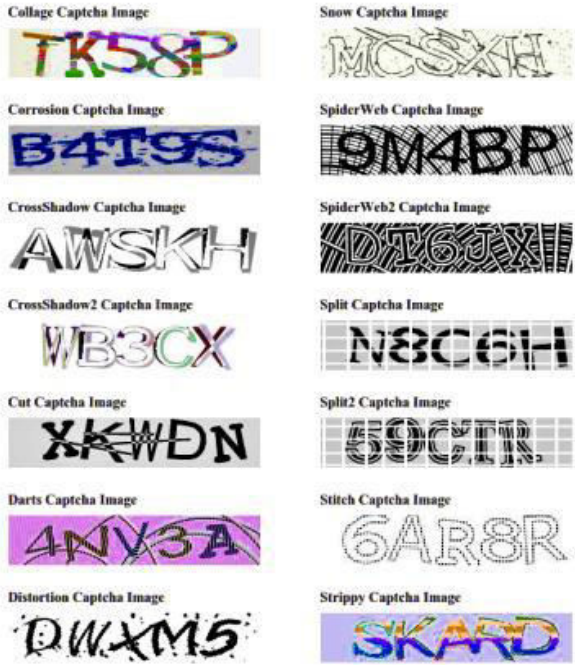

Fig. 1: Examples of different alphanumerical CAPTCHAs

This research aims to crack the CAPTCHAs high vulnerability problem, to identify basic weaknesses and vulnerabilities, and improve CAPTCHA generation technology so that it does not lag behind the ever-increasing intelligence of bots and scams.

## II. RELATED WORK

Since many websites actively utilize CAPTCHAs to protect traffic, many organizations have already invested significant resources in breaking CAPTCHAs to evaluate these data techniques' strengths and weaknesses.

Segmentation-based systems to analyze CAPTCHAs are still prevalently used for CAPTCHA-detection. The segmentation step is the most crucial component of these segmentation-based models' recognition process. Several techniques for segmenting text-based CAPTCHAs into different characters have been proposed. However, with recent CAPTCHAs, single characters cannot be effectively divided using rectangular windows as they may overlap. These CAPTCHAs resemble handwritten text and can be solved with a CNN architectural model. Researchers have recently begun implementing segmentation-free models to determine CAPTCHAs directly without segmentation to overcome the drawbacks of poor CAPTCHA segmentation methods.

## III. PROPOSED METHOD

To crack the CAPTCHA, this research displays a deep neural network architecture named CapNet. Since the training images represent a 10000 - dimensional (200-by-50 pixels, pre-processed into grayscale) feature space, we flatten the vector and make five branches from it, each branch predicting one letter, using TensorFlow as the back end and Keras as the front-end to train our model by implementing customized convolutional layers to fit our needs by adjusting various parameters like the total number of parameters, layers, activations, filter sizes, loss functions, and more as per our needs. The exact approach for processing, identifying, and cracking the alphanumerical CAPTCHA images is described further below. The process includes pre-processing input data, output encoding, and the network structure.

In addition to developing the neural net architecture, transfer learning was also examined with a pre-trained neural network on ImageNet. The specific net that was utilized is named VGG-19, which has over 20 million free parameters (shown in Fig.3). Several of the last convolutional layers were deliberately froze while maintaining the initial few since they are capable of classifying high-level features necessary for pattern recognition.

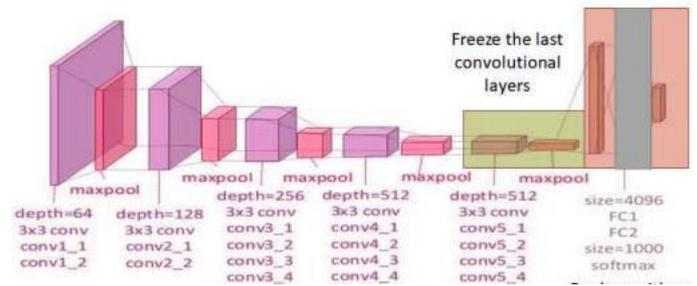

Fig.2 Structure of VGG-19 and our freezing of many last convolutional layers

In addition to this several other pre-existing models like Alexnet, have also been used, most of which returned a low testing accuracy shown in fig 3.

| VGG-16 | AlexNet | CapNet |
|---|---|---|
| Training ---> 98.4% | Training ---> 95% | Training---> 96% |
| Testing ----> 87% | Testing ---> 83% | Testing---> 94.67 |

Fig.3 Comparison of various other model's results with CapNet

Out of which the least overfitting was produced by the proposed model CapNet (fig 3.2) with a training accuracy of 96% and 94.67% test.

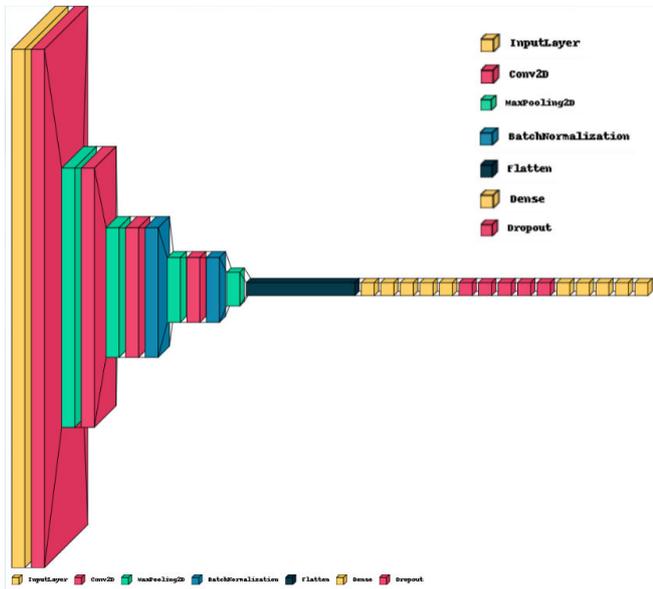

Fig.3.2 Comparison of various other model's results with CapNet

*4.1 Dataset Generation*

Since there are no hand-labeled CAPTCHA datasets available, the dataset was generated using a python library (captcha 0.4) while ensuring that there are no duplicates in the dataset. The library as mentioned above, is made up of distorted alphanumeric characters with a constant length of 5. They are 50x200 in size. We also avoid using dictionary words but instead use random characters, removing the requirement that every second character be a vowel. Fig 4 shows some examples of the auto generated captchas.

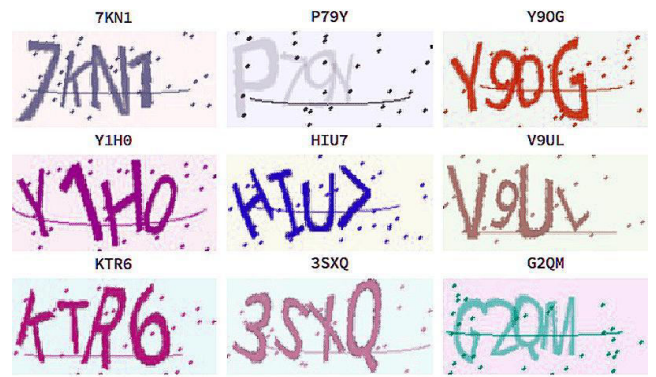

Fig.4 sample of auto generated CAPTCHAs using python library

*4.2 Preprocessing*

Applying pre-processing techniques like noise reduction filtering, Grey-Scaling, Resizing, Normalization, One-Hot Encoding, and image size reduction can significantly boost network performance. Gray-scale conversion is another preprocessing method that can be used to reduce the size of the data while maintaining a similar level of accuracy in detection, using this we can minimize the amount of redundant data even further and ease the model training and prediction. Converting a three-channel RGB image to a greyscale image has no effect on the results because color isn't important in text-based CAPTCHA systems.

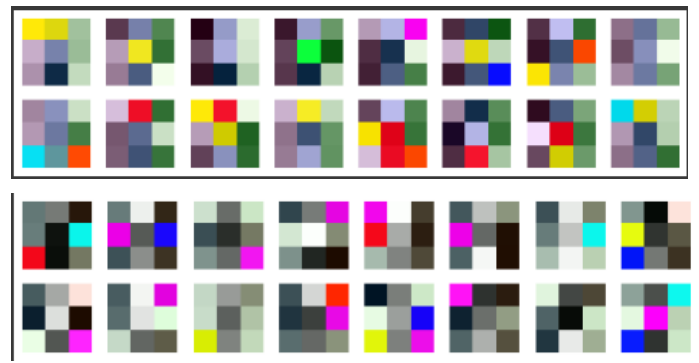

Fig.5.1 example of filters

*4.3 Architecture of the Model*

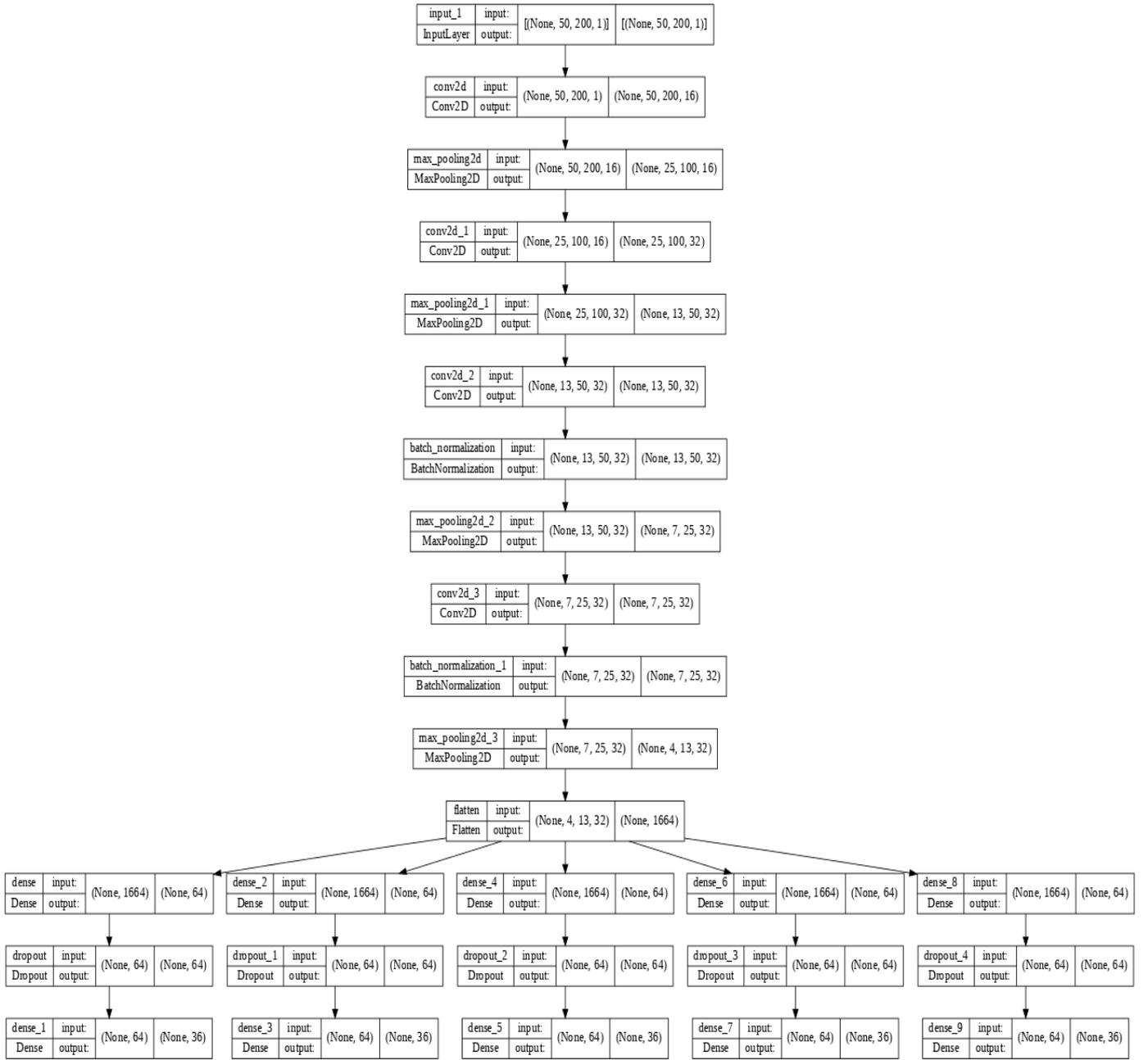

Fig.6 Architecture of proposed model

The proposed architecture begins with a Convolutional layer, the ReLU activation function, and 3x3 Kernels followed by a Max-Pooling layer. Then we have another Convolutional-MaxPooling pair with the same parameters except for the number of neurons, which is set by two sets of Convolutional-Batch Normalization-MaxPooling layers. After the Convolutional layers, a Flatten layer prior to five separate 1664-dense layer with the ReLU activation function and a drop-out layer follows. Finally, we have five distinct Softmax layers. Therefore, to compare these binary matrices, the suggested network's loss function is the Binary-cross entropy:

$$H_p(q) = -\frac{1}{N} \sum_{i=1}^{N} y_i \cdot \log(p(x_i)) + (1 - y_i) \cdot \log(1 - p(x_i))$$

where N is the sample count, and p is the predictor. Values xi and yi denote the input data and the label of the I th sample. As the given label can be either 0 or 1, only a part of this equation will remain active for each sample. Adam optimizer was also used, which is briefly explained in the following equations, with mt and vt denoting a decaying average of previous gradients and past squared gradients, respectively

$$m_t = \beta_1 m_{t-1} + (1 - \beta_1)g_t$$
$$v_t = \beta_2 v_{t-1} + (1 - \beta_2)g_t^2$$

where b1 and b2 are the configurable constants, gt - gradient of the optimizing function followed by t - learning iteration.

$$\hat{m}_t = \frac{m_t}{1 - \beta_1^t}$$

$$\hat{v}_t = \frac{v_t}{1 - \beta_2^t}$$

Finally, the value of the function can be obtained by using the following equation and updating t in each iteration. In our approach, m t and vt are determined using the above equations, and the learning rate is set at 0.001.

$$\theta_{t+1} = \theta_t - \frac{\eta}{\sqrt{\hat{v}_t} + \epsilon}\hat{m}_t$$

The logic behind using the Adam optimizer is its ability to train the network in a decent amount of time as seen above, this optimizer achieves similar results as other optimizers but with substantially faster convergence.

The network was trained for 200 epochs with a batch size of 32 for each after multiple experiments. Figure 7 shows that even after 100 epochs, the network tends to an acceptable convergence. Consequently, 200 epochs appear to be enough for the network to execute consistently. Furthermore, Figure 7 implies the same conclusion based on the metrics of the measured accuracy.

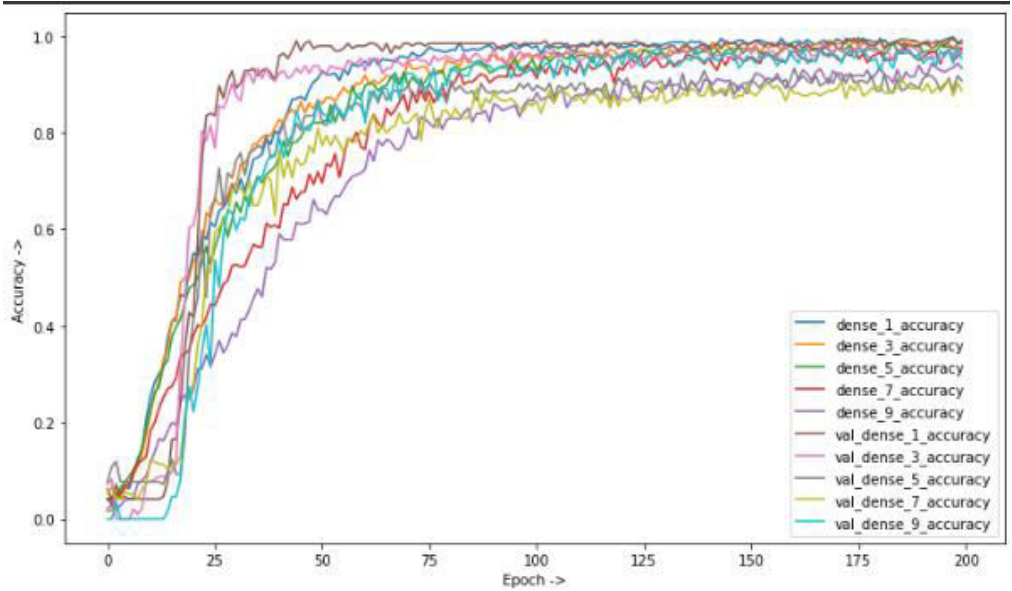

Fig.7 Accuracy vs Epochs Plot

*5.1 Evaluation*

After training CapNet for 200 epochs. The total training time results to be ~32000 ms with an avg epoch time of ~160 ms and a batch size of 32 the model reached total performance with an accuracy rate of ~96.5% on the training set and ~96% on testing. The obtained accuracy of CapNet can be calculated using two criterias : Character recognition accuracy and total CAPTCHA prediction accuracy. All the characters of a CAPTCHA are classified individually for absolute character recognition accuracy which is further calculated by dividing the sum of correctly predicted characters by the total number of present characters. For total CAPTCHA prediction accuracy, all the characters of the CAPTCHA must be classified correctly; if any one of the five characters of a CAPTCHA is incorrectly classified, the recognition results to incorrect recognition.

Since obtaining training samples is the most expensive part, this approach seeks to reduce the size of the initial training set to less than 1000 images.

| Digits | Training loss | Testing loss |
|---|---|---|
| Digit 1 | 0.045 | 0.012 |
| Digit 2 | 0.006 | 0.015 |
| DIgit 3 | 0.009 | 0.029 |
| Digit 4 | 0.009 | 0.038 |
| Digit 5 | 0.010 | 0.035 |
| CapNet | 0.015 | 0.025 |

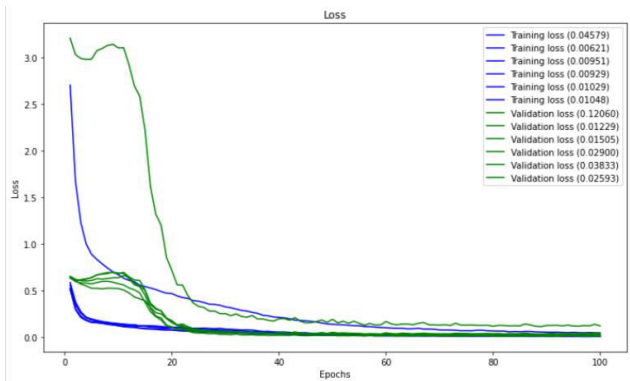

Fig.8.1 Loss vs Epochs Plot

| Digits | Training Accuracy | Testing Accuracy |
|--------|-------------------|------------------|
| Digit 1 | 98.19 % | 98.45 % |
| Digit 2 | 96.00 % | 98.45 % |
| DIgit 3 | 97.03 % | 94.84 % |
| Digit 4 | 95.74 % | 94.33 % |
| Digit 5 | 95.74 % | 94.33 % |
| CapNet | 96.54 % | 96.08 % |

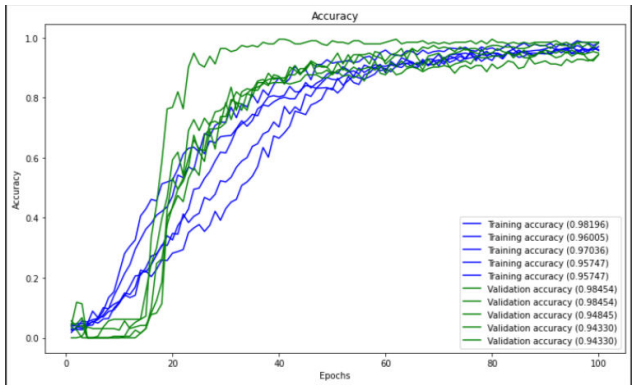

Fig.8.2 Accuracy vs Epochs Plot

Emphasizing that the above metrics of accuracy are determined by the total number of CAPTCHAs that are correctly detected, else individual digit accuracy is even higher.

After experimenting with several other loss functions, Binary-cross-entropy yielded the best results as expressed in fig 9.

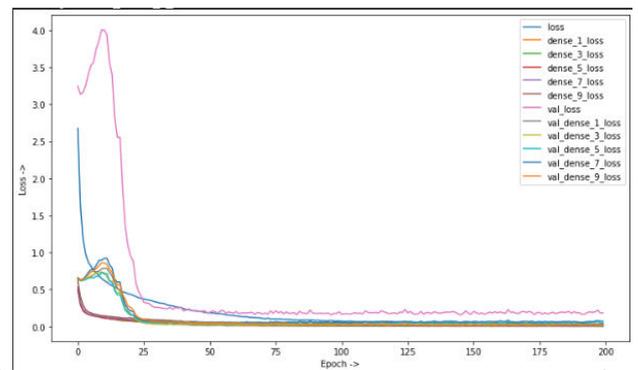

Fig.9 Binary cross entropy Vs Stochastic Gradient Descent

### 5.2 Vulnerability Analysis

We estimate the predictive distribution of each digit by normalizing the sum of the network outputs to 1, then calculate the overall uncertainty using "best-vs-second-best" as a criterion.

$$\eta = \frac{1}{d} \cdot \sum_{i=1}^{d} \frac{\arg\max \{\mathcal{P}(x_i) \setminus \arg\max \mathcal{P}(x_i)\}}{\arg\max \mathcal{P}(x_i)},$$

While an average person can solve most of the wonrlgy classified CAPTCHAs, these flaws caused CapNet to underperform:

- The intensity of gray-level in the generated CAPTCHAs was significantly lower compared to the average of the Gaussian distributed pepper noise in the CAPTCHAs in majority of misclassified samples.
- Most commonly misclassified digits were 3, 8 & 9.
- Rotation of the character resulted in more misclassification.

A similar analysis was carried out for the alphabets of the failed CapNet detections. Most of the unsuccessful predictions were related to either too-orientated characters or those in immediate contact with adjacent letters. As an example, the letter "g" may be misclassified as "9" in certain angles, or a "w" may be misidentified as "m" when it comes into contact with an upright letter like "T." As a result, those letters which can be associated with one or more of the following letters: w, v, m, n can create difficulty for CapNet to predict these images.

As a result, this research proposes that including more of these letters and placing them in close proximity to other letters lowers the vulnerability of the CAPTCHAs. This research also reveals that alpha-numerical characters with brighter color with low grayscale intensity will aid in increasing the difficulty level of the CAPTCHAs solving systems.

## IV. CONCLUSION

This research presents a CAPTCHA-solving algorithm that trains a deep CNN with a small dataset to reduce time and computation while providing high accuracy to highlight the strengths and drawbacks of standard CAPTCHA generators.

The experimental results suggest that the proposed method outperforms other models that use relatively smaller datasets in terms of CAPTCHA recognition accuracy. The proposed model also has a substantially smaller storage size than the different versions.

There are numerous opportunities for future work. The most straightforward approach would be to extend the initial training set with large amounts of data. This is to see if we can approximate all the additional font styles that might be used to construct a CAPTCHA without specifically incorporating them during training.

Additionally, hyperparameter optimization could be used to try to optimize the unique architecture (which also requires significantly more computing power). However, because the parameter space for CNNs is so large, other approaches, like as metamodeling, may be more viable.